\documentclass[letterpaper,fleqn,usenatbib]{mnras}

\usepackage{newtxtext,newtxmath}

\usepackage[T1]{fontenc}
\usepackage{ae,aecompl}

\usepackage{graphicx}	
\usepackage{amsmath}	
\usepackage{amssymb}	

\title[Cooling in shadowed rings]{Cooling in the shade of warped transition disks}  

\author[S. Casassus et al.]{
Simon Casassus,$^{1,2}$\thanks{E-mail: simon@das.uchile.cl}
Sebasti\'an P\'erez,$^{1,2,3}$
Axel Osses,$^{4}$
Sebasti\'an Marino,$^{5}$
\\
$^{1}$Departamento de Astronom\'{\i}a, Universidad de Chile, Casilla 36-D, Santiago, Chile\\
$^{2}$Millennium Nucleus ``Protoplanetary Disks'', Chile\\
$^{3}$Universidad de Santiago de Chile, Av. Ecuador 3659, Santiago\\
$^{4}$Departamento de Ingenier\'{\i}a Matem\'atica, Facultad de Ciencias F\'{\i}sicas y Matem\'aticas, Universidad de Chile, Beauchef 851, Santiago, Chile \\
$^{5}$Max Planck Institute for Astronomy, K\"onigstuhl 17, 69117 Heidelberg, Germany\\
}

\date{Accepted XXX. Received YYY; in original form ZZZ}

\pubyear{2019}

\begin{document}
\label{firstpage}
\pagerange{\pageref{firstpage}--\pageref{lastpage}}
\maketitle

\begin{abstract}
    The mass of the gaseous reservoir in young circumstellar disks is
    a crucial initial condition for the formation of planetary
    systems, but estimates vary by orders of magnitude. In some disks
    with resolvable cavities, sharp inner disk warps cast two-sided
    shadows on the outer rings; can the cooling of the gas as it
    crosses the shadows bring constraints on its mass?  The finite
    cooling timescale should result in dust temperature decrements
    shifted ahead of the optical/IR shadows in the direction of
    rotation.  However, some systems show temperature drops, while
    others do not. The depth of the drops and the amplitude of the
    shift depend on the outer disk surface density $\Sigma$ through
    the extent of cooling during the shadow crossing time, and also on
    the efficiency of radiative diffusion. These phenomena may bear
    observational counterparts, which we describe with a simple
    one-dimensional model. An application to the HD\,142527 disk
    suggests an asymmetry in its shadows, and predicts a
    $\gtrsim$10\,deg shift for a massive gaseous disk, with peak
    $\Sigma > 8.3$\,g\,cm$^{-2}$. Another application to the DoAr~44
    disk limits the peak surface density to $\Sigma <
    13$\,g\,cm$^{-2}$.
\end{abstract}





\section{Introduction}

Warped disk geometries have been proposed to account for optical/IR
(OIR) illumination effects seen in high-contrast images of Class\,{\sc
  ii} young stellar objects. Such two-sided shadows stand out in the
outer disks of systems with a resolvable central cavity, i.e. in
transition disks, as in HD\,142527 \citep{Marino2015ApJ...798L..44M},
HD\,100453 \citep[][]{Casassus2016PASA...33...13C,
  Benisty2017A&A...597A..42B, Long2017ApJ...838...62L}, DoAr\,44
\citep[][]{Casassus2018MNRAS.477.5104C}, and HD\,143006
\citep[][]{Benisty2018arXiv180901082B}.  The position and shape of the
shadows place constraints on the orientation of the inner disk
relative to the outer disk \citep{Min2017A&A...604L..10M}, with
consistent kinematics when resolved CO or HCO$^+$ rotational line data
are available \citep[][]{Casassus2015ApJ...811...92C,
  PerezL2018ApJ...869L..50P}.


An interesting consequence of tilted inner disks is that the outer
rings are directly exposed to stellar radiations, except in the
shadowed regions. If the shadow crossing time is larger or comparable
to the gas cooling timescale, if $\Delta t_K \gtrsim \Delta t_c$, and
depending on the extent of radiation diffusion from the rest of the
disk, the gas and dust under the shadows may cool down appreciably,
possibly resulting in a radio continuum drop. This seems to be the
case in HD~142527, with a temperature drop of $\sim$30\% in its
northern shadow, as inferred from the grey-body analysis in
\citet[][]{Casassus2015ApJ...812..126C}. Intriguingly, the existing
$^{12}$CO data in HD\,142527 do not show any decrement, and the
southern shadow is not matched by a continuum decrement as clear as in
the northern shadow.  Radio continuum counterparts to the OIR
decrements have also been seen in DoAr\,44
\citep{Casassus2018MNRAS.477.5104C}, and in J1604-2130
\citep[][]{Mayama2018ApJ...868L...3M}, even though in this source the
shadows are variable \citep[][]{Pinilla2018ApJ...868...85P}. However,
they are absent in other warped disks, such as in HD\,143006
\citep[][]{PerezL2018ApJ...869L..50P} and in HD\,100453
\citep[][]{2019arXiv190200720V}. Why is there such a broad variety in
radio continuum responses under the shadows?

If $\Delta t_K \sim \Delta t_c$, the thermal lag in the shadowed gas
and dust may be important, and the radio decrements could also lead
the OIR shadows in the direction of Keplerian rotation, by an angular
shift $\eta_S$. Can the position and shape of the temperature
decrements constrain the outer disk mass?  The depth of the radio
decrements and the shift $\eta_S$ may provide a measurement of the gas
cooling timescale, which depends on the internal energy reservoir and
hence on the gas mass. Quantitative predictions and limits based on
measurement accuracies will require full-blown radiative transfer (RT)
models that include the advection of internal energy in a prescribed
velocity field. Meanwhile, Section~\ref{sec:tcool} presents a
one-dimensional model for the azimuthal temperature profile in the
shadowed material, which Sec.~\ref{sec:applications} applies to the
cases of HD\,142527 and DoAr\,44. Sec.~\ref{sec:conclusion} concludes
on the two questions that motivate this letter.

\section{Model for the temperature drop of shadowed material}  \label{sec:tcool} 

In a two-layer model for a passive disk
\citep{ChiangGoldreich1997ApJ...490..368C}, where the bulk mass is in
the interior at a temperature $T_g$, the internal energy per unit
surface in the disk is 
\begin{equation}
U \approx \frac{1}{\gamma-1} k \frac{\Sigma}{\mu m_p} T_g,   \label{Eq:U}
\end{equation}
where $\Sigma$ is the gas mass surface density, $\gamma$ is the
adiabatic index ($\gamma \approx 1.4$ for diatomic gas), $k$ is the
Boltzmann constant, $\mu=2.3$ is the mean molecular weight and $m_p$
is the proton mass. An approximation to cooling per unit area in the
disk due to radiation from both disk surfaces is given by dust
emission at a representative temperature $T_d$,
\begin{equation}
  F_c \approx - 2 (1-\exp(-\tau_P)) \sigma T_d^4, \label{Eq:F_c} 
\end{equation}
where $\tau_P=\Sigma \kappa_P$, and $\kappa_P$ is the Planck-averaged
absorption opacity \citep[as in the deviation coefficients used
  by][]{Dullemond2001ApJ...560..957D}. In the absence of shadows, the
heating per unit area of the interior layer due to the stellar
radiation transferred by the two surface layers can be written in terms
of another temperature $T_\circ$,
\begin{equation}
F_h \approx 2 (1-\exp(-\tau_P)) \sigma T_\circ^4. \label{Eq:F_h}
\end{equation}
The factor  $2(1-\exp(-\tau_P))$ in Eq.\,\ref{Eq:F_h} allows for
$T_d = T_\circ$ in steady state, and in the absence of shadows,
i.e. $T_\circ$ is the equilibrium temperature of the irradiated disk,
as in \citet{ChiangGoldreich1997ApJ...490..368C}.




We will assume perfect thermal coupling of dust and gas, so that $T_g
= T_d= T$ for the interior layer. The error involved in this
approximation is within $\sim 20\%$, if it is similar to the
difference in equilibrium temperature, evaluated at the midplane, for
the small and large grain populations in the RT models used in
Sec.\,\ref{sec:applications}. Under the isothermal approximation and
with perfect thermal coupling, the cooling timescale can be defined as
\begin{equation}
  \Delta t_c \equiv \left| \frac{U}{\dot{U}} \right| =    \frac{1}{\gamma-1} \frac{k \Sigma}{ 2 (1-\exp(-\tau_P)) \sigma  \mu m_p T_\circ^3}. \label{eq:Delta_tc}
\end{equation}
If $T_\circ$ drops suddenly to $0$ at $\phi=0$ when entering a shadow
that extends to $+\infty$ along a straight line $\vec{s}=R\phi\hat{s}$, then
(keeping $\tau_P$ fixed at the unshadowed value),
\begin{equation}
T \approx T_\circ \left[ 1+ \frac{3 R \phi}{v \Delta t_c}
  \right]^{-1/3}. \label{eq:nodiffusion}
\end{equation}
This approximation neglects cooling due to transport of mechanical
energy, as in the work required in launching spiral arms
\citep[][]{Montesinos2016ApJ...823L...8M}. We assume that the power
dissipated in spirals is similar to viscous heat dissipation, which is
much less than that due to stellar radiation beyond a few au
\citep[e.g.][]{Bitsch2013A&A...549A.124B}.







In optically thick media, with a gas density $\rho$, the transport of
radiative energy density is due to diffusion, with a local flux
\begin{equation}
g_D = - \Lambda \frac{16 \sigma T^3}{3\kappa_R \rho}
\frac{dT}{ds},  \label{eq:F_D_local}
\end{equation}
in direction $\hat{s}$. The Rosseland-mean opacity
$\kappa_R(T)$ is given by
\begin{equation}
\frac{1}{\kappa_R(T)} = \frac{\int_0^\infty d\nu (1/\kappa_\nu) \partial B_\nu(T)  / \partial T}{\int_0^\infty d\nu  \partial B_\nu(T)  / \partial T},
\end{equation}
in which $\kappa_\nu$ is the total mass opacity at frequency
$\nu$. The dimensionless control factor $\Lambda=1$ for the Rosseland
approximation, which requires a small photon effective mean-free-path
\citep[the net displacement due to scattering before
  absorption,][]{Rybicki1986rpa..book.....R}, i.e. $l_\star \ll H$ at the
Wien wavelength. This condition can be approximated as $\tau_R 
\equiv \kappa_R \Sigma \gg 1$. In the applications below $ 0.01< \tau_R
< 27 $, so in order to extend Eq.\,\ref{eq:F_D_local} to the optically
thin transition we parametrise the control factor as in \citet[][their
  Eqs.\,11 and 12]{Dobbs-Dixon2010ApJ...710.1395D}.



\begin{figure}
\begin{center}
  \includegraphics[width=\columnwidth,height=!]{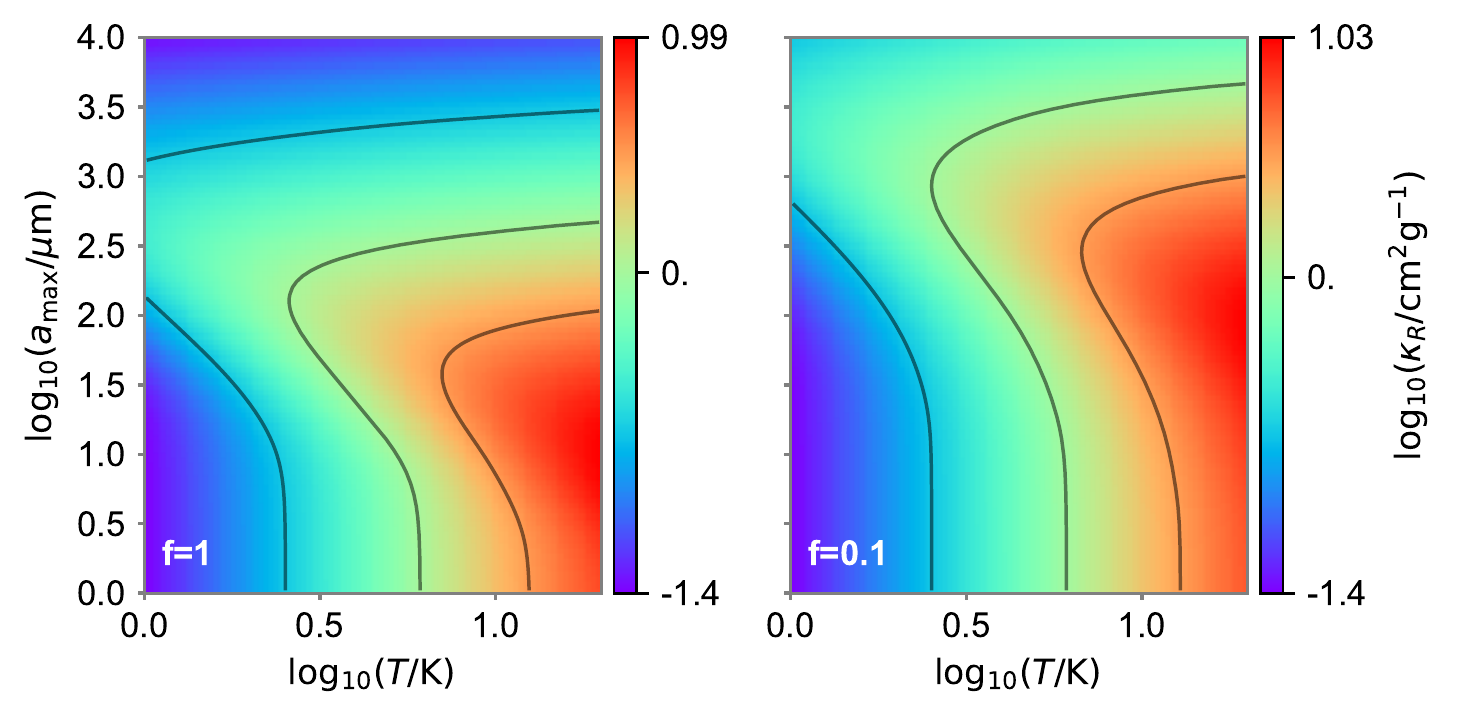}
\end{center}
\caption{$\log_{10}(\kappa_R(a_\mathrm{max},T))$ for two dust filling
  factors, $f=1$ and $f=0.1$. The 3 contours correspond to half the
  extrema of the colour bar and 0.
  \label{fig:kR}}
\end{figure}

The radii of cavities that can be resolved by direct imaging are
$\gtrsim$30\,au, where the equilibrium temperature in the midplane is
usually $<100$\,K for Herbig Ae/Be and TTauri stars. The mass opacity
$\kappa_\nu$ is therefore due to grains with water ice mantles. The
available formulae for $\kappa_R(T)$ focus on higher temperatures and
use a small maximum grain size, typically of $a_\mathrm{max} \sim
1\,\mu$m \citep[e.g.][]{Zhu2012ApJ...746..110Z}.  Since the large
rings in Class\,{\sc ii} disks are associated with dust trapping, and
possibly grain growth, we compute $\kappa_R(T,a_\mathrm{max})$ for a
mix of water ice, silicates and graphite, with mass fractions of 0.4,
0.3, and 0.3, and an internal grain density $\rho_\bullet =
1.8$\,g\,cm$^{-3}$. The analytical formulae in
\citet{Kataoka2014A&A...568A..42K} give the absorption and scattering
cross sections $Q^\mathrm{abs}_\nu(a)$ and $Q^\mathrm{sca}_\nu(a)$ for
spherical grains with radii $a$ and filling factor $f$. For a
power-law size distribution with exponent $q$,
\begin{equation}
  \kappa_\nu = \frac{1}{f_\mathrm{gd}} \frac{3}{4 f \rho_\bullet}
  \frac{q+4}{a^{q+4}_\mathrm{max} - a^{q+4}_\mathrm{min}}
    \int_{a_\mathrm{min}}^{a_\mathrm{max}} da 
    (Q^\mathrm{abs}_\nu+Q^\mathrm{sca}_\nu) a^{q+2}.
\end{equation}
We set $q=-3.0$; this rather shallow value anticipates an application
to transition disk rings, thought to undergo grain growth. The
resulting Rosseland-mean opacities $\kappa_R(a_\mathrm{max},T)$ are
shown in Fig.\,\ref{fig:kR}, for a gas-to-dust mass ratio
$f_\mathrm{gd}=100$.


The total radiative flux in the azimuthal direction, i.e. in the
direction $\hat{\phi}$, through a section of the
disk at constant $\phi$, is
\begin{equation}
\mathcal{F}^D_\phi = \int dr \int dz g_D ~ \approx \Delta R \int dz
g_D, \label{eq:F_D}
\end{equation}
for some small interval in stellocentric radius $\Delta R / R \ll
1$. Eq.\,\ref{eq:F_D_local} is applicable only to the disk interior,
above the disk surface the radiation field points in the vertical
direction, whereas in the low density limit Eq.\,\ref{eq:F_D_local}
yields equal components in all directions.  We  assume that
azimuthal radiation transport occurs at mid-plane densities, and that
it extends over a vertical full-width-half-maximum,
\begin{equation}
\mathcal{F}^D_\phi \approx \Delta R  \sqrt{2\pi} h R   g_D|_\mathrm{midplane}, \label{eq:fluxapprox}
\end{equation}
for a hydrostatic and vertically isothermal disk with $1\sigma$
thickness $H = h R$, where $h$ is the $1\sigma$ disk aspect ratio. We
checked this approximation {\em a-posteriori}, using the temperature
profiles in the applications of Sec.\,\ref{sec:applications} by
modulating Eq.\,\ref{eq:F_D_local} with the hydrostatic Gaussian,
finding that Eq.\,\ref{eq:fluxapprox} underestimates Eq.\ref{eq:F_D}
by a factor $\sim 3\pm1$, which we chose to ignore.  The net heating
per unit disk surface is thus
\begin{equation}
  F^D_\phi = - \frac{1}{\Delta R} \frac{\partial \mathcal{F}^D_\phi}{R\partial \phi}  \approx  \frac{\partial }{R\partial \phi}  2\pi H^2 \Lambda  \frac{16 \sigma T^3}{3 \kappa_R \Sigma}  \frac{\partial T }{R\partial \phi}. \label{eq:F^D_phi}  
\end{equation}
Likewise, in the radial direction, the total flux of radiative
diffusion through an arc length $R\Delta \phi$ is
$\mathcal{F}^D_R \approx R \Delta \phi  \sqrt{2\pi} H   g_D|_\mathrm{midplane}$, and the net heating  per unit disk surface is 
\begin{equation}
  F^D_R = -  \frac{1}{R\Delta \phi} \frac{\partial \mathcal{F}^D_R
  }{\partial R} =   \frac{1}{R} \frac{\partial }{\partial R} R 2\pi H^2  \Lambda \frac{16 \sigma T^3}{3 \kappa_R \Sigma}  \frac{\partial T }{\partial R}. \label{eq:F^D_R}  
\end{equation}
Radial diffusion turns out to be small compared to $F_c$. For a
power-law disk with $H \propto R^{-\beta + \frac{3}{2}}$, $T\propto
R^{-2\beta}$, $\Sigma \propto R ^{-\gamma}$, a constant $\kappa_R$,
$\beta = \frac{3}{8}$, setting $\Lambda=1$ and at the peak surface
density with $\gamma = 0$,
\begin{equation}
  F^D_R =  \frac{3 \pi} {\kappa_R \Sigma}   \frac{H^2}{R^2}  2 \sigma  T^4 ~     \approx - 0.1 F_c,
\end{equation}
where the last approximation assumes $\kappa_R \Sigma \sim 1$ and
$h\sim 0.1$.

For an incompressible Keplerian flow, at a fixed radius $R$ and at
constant velocity $v$ in the direction $\hat{s} \parallel \hat{\phi}$,
the net internal energy advected into a volume $d \mathcal V =
\sqrt{2\pi} H R dR d\phi$ is $ -v dR d\phi \, \frac{\partial U}{
  \partial \phi}$. The net heating per unit disk surface due to
advection is thus
\begin{equation}
F_a =  -\frac{v}{R}\, \frac{\partial U}{ \partial \phi},
\end{equation}
This advection term is similar to that in the usual convective
derivative. We neglect the random-walk delay associated to vertical
diffusion, $ \tau_\nu \times H / c$, which is of order a few days for
IR photons.

Energy balance requires that
\begin{equation}
F_a   +   F^D_\phi  +  F_c     +   F_h   =    \frac{\partial U}{\partial t}, \label{eq:diffEq0}
\end{equation}
where we have neglected radial diffusion. Following Eq.\,\ref{Eq:F_h},
we parametrise the heat source with an effective temperature $T_h$, so
that $F_h(\phi) = 2 (1-\exp(-\tau_R)) \sigma T^4_h(\phi)$.  We replace
$\tau_P$ by $\tau_R$ for simplicity, and Eq.\,\ref{eq:diffEq0} gives: 
\begin{multline}
  -\frac{1}{\gamma-1}  v  \Sigma \frac{k}{\mu m_p} \frac{\partial T}{ \partial \phi}
  + 2\pi H^2 \frac{16 \sigma   } {3 \Sigma R} \frac{\partial }{\partial \phi} \left( \frac{f_\mathrm{gd}}{100} \frac{\Lambda \, T^3}{\kappa_R(T,a_\mathrm{max}) } \frac{\partial T}{\partial \phi} \right) \\
  - 2\sigma (1-\exp(-\tau_R)) R (T^4  - T_h^4 ) =  R \frac{\partial U}{\partial t}. \label{eq:mastereq}
\end{multline}
In steady state $\frac{ \partial U}{\partial t }=0$. We solve
Eq.~\ref{eq:mastereq} with centered second-order finite differences,
starting with a flat initial condition, and with periodic boundary conditions
in $\phi \in [0,2\pi]$.




We can now estimate a spatial scale associated to the smoothing effect
of azimuthal radiation transport. For an infinite line along
$s=R\phi$, in the absence of a heat source beyond the onset of a
shadow at $s=0$, without advection, and in the limit $\tau_R \gg 1$,
Eq.\,\ref{eq:mastereq} yields an exponential decay with e-folding
length $ L_D/H \approx \sqrt{f_\mathrm{gd}/(100 \tau_R)} \ll 1$.
However, if $\tau_R \ll 1$, assuming exponential decay yields $ L_D/H
\approx f_\mathrm{gd}/(100 \tau_R)$, and radiation will strongly
smooth out any temperature decrement. Likewise, away from the
mid-plane, Eq.\,\ref{eq:F_D_local} suggests larger radiation flux at
lower $\rho$. In the rarefied disk surface, e.g. where $^{12}$CO
rotational lines originate, the medium is optically thin and we 
qualitatively expect  that the much larger photon mean-free path will smooth
out the decrements.





\section{Applications} \label{sec:applications}

The shape of the shadows and $T_h(\phi)$ can potentially be
constrained by observations and RT modeling. If we consider the
details of $T_h(\phi)$ as fixed, as well as $R$ and $H$, which are
given by the disk structure, then in this simplified model the only
free parameters governing the gas temperature as it crosses the
shadows are: the disk surface density $\Sigma$, the gas-to-dust mass
ratio $f_\mathrm{gd}$, the maximum grain-size $a_\mathrm{max}$, and
the grain filling factor $f$.






\subsection{HD\,142527}






The very lopsided ring of HD\,142527 \citep[][]{Casassus2013Natur}
originates fairly optically thick mm-continuum in a stratified disk,
so that the continuum profile is not axially symmetric, which hampers
an isothermal description.  A detailed comparison of the effect of
advection under the shadows requires 3D RT models, as well as fine
angular resolution data to infer the observed continuum temperature
profile. Also, despite this source being the first in which sharp
shadows from an inner disk were inferred, the precise location of the
center of the shadows in the outer disk midplane has not yet been
worked out. For now we use our simplified model to make qualitative
predictions on the shape of the temperature decrements.

\begin{figure}
\begin{center}
  \includegraphics[width=\columnwidth,height=!]{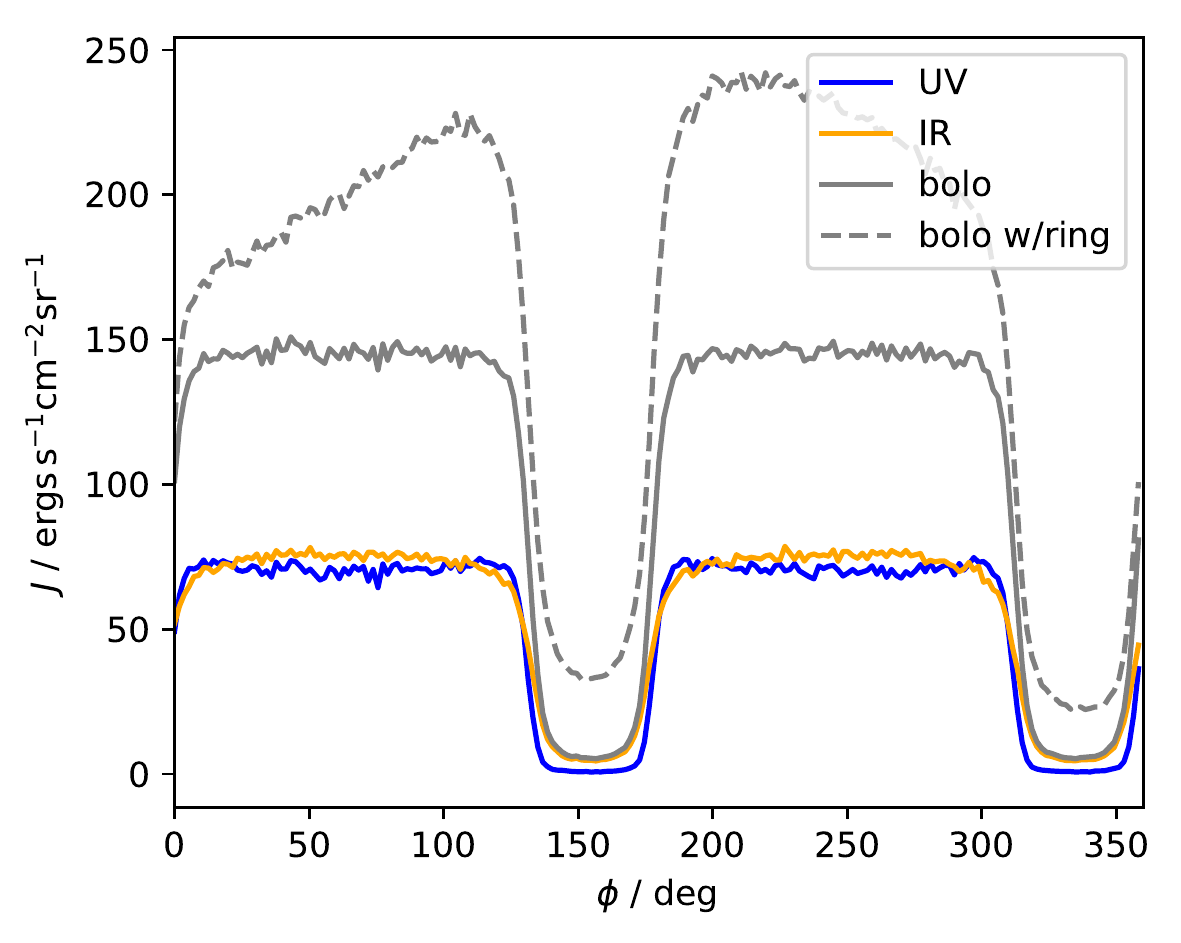}
\end{center}
\caption{ Mean  intensity profile $J(\phi) = \int d\nu
  J_\nu(\phi)$, as a function of azimuth $\phi$, at the outer edge of
  the cavity and in the midplane of a parametric model for
  HD\,142527. We compare the UV/optical field, $J_\mathrm{UV}$
  corresponding to $\lambda < 1\mu$m, with the IR field,
  $J_\mathrm{IR}$, for $\lambda>1\mu$m, and with the bolometric field
  $J_\mathrm{bolo}$. All profiles are calculated in a model without
  outer disk, except for the dashed line, labeled `w/ring', which
  corresponds to $J^\mathrm{w/ring}_\mathrm{bolo}$.  \label{fig:J}}.
\end{figure}

Under the shadows some radiative heating persists mainly due to IR
thermal radiation from the inner disk. To guide the choice of the heat
source $T_h(\phi)$, we examine the mean radiation intensity field in a
parametric RT model meant to approximate the disk of HD\,142527 at a
distance of 140\,pc, so somewhat less than the GAIA distance of
156\,pc \citep[hereafter the RT$_\mathrm{1425}$
  model,][]{Marino2015ApJ...798L..44M, Casassus2015ApJ...811...92C,
  Casassus2015ApJ...812..126C}. This model was computed with the
RADMC3D package \citep[][]{RADMC3D0.41}. The gas and the small grains
are heated by UV radiation in the surface layers, while dust in the
midplane will absorb transferred IR radiation. Using the
RT$_\mathrm{1425}$ model, we compute the mean radiation intensity
$J(\phi)$ inside the cavity, right at the inner edge and in the
midplane of the gaseous outer disk (so at a radius of 115\,au).
Fig.\,\ref{fig:J} compares the mean intensity field for $\lambda <
1\,\mu$m, $J_\mathrm{UV}$, with the mean field for $\lambda >
1\,\mu$m, $J_\mathrm{IR}$, and with the bolometric field
$J_\mathrm{bolo}$, both in the presence and absence of the outer
disk. About one third of $J^\mathrm{w/ring}_\mathrm{bolo}$ is thermal
emission from the outer disk, hence its harmonic modulation in azimuth
and its smoother profile under the shadows.



Since the dust responsible for the mm-continuum is thought to be
trapped at the pressure maximum in the outer disk, radiation will be
attenuated as it is transferred through the intervening material. This
can be incorporated by setting the unshadowed temperature $T_\circ$ to
the temperature of the larger grains (i.e.  1\,mm -- 1\,cm).  Since
Eq.\,\ref{eq:mastereq} accounts for outer disk radiation in the
limited diffusion approximation, once we have set $T_\circ$,
$T_h(\phi)$ should follow only radiation stemming from the star and
the inner disk, which appears to be fairly square.  The choice of a
square profile for $T_h(\phi)$, with a total width $\phi_S$ for each
shadow, and a floor temperature $T_S$, is also useful to estimate how
the shape of the shadows maps into the temperature profile
$T(\phi)$. We use hyperbolic tangents to round  the edges.




The temperature for the largest grains close to the northern shadow is
observed to be $T_\circ \sim 27.5\,$K
\citep[][]{Casassus2015ApJ...812..126C}. The floor temperature $T_S$
can be inferred from the drop in radiation intensity, from $J_\circ$
in the unshadowed regions, to $J_S$ under the shadows, $ \frac{T_S}{
  T_\circ} \sim (J_S/J_\circ)^{1/4}$.  Inspired by
$J_\mathrm{bolo}(\phi)$ in Fig.~\ref{fig:J}, we set $T_S =
0.44\,T_\circ$ for RT$_\mathrm{1425}$.


%




The peak gas surface density in the RT$_\mathrm{1425}$ model is
$\Sigma \sim$83\,g\,cm$^{-2}$, at $R = $155\,au, for a gas to dust
mass ratio $f_\mathrm{gd}=100$. The corresponding peak dust surface
density is similar, within 50\%, to that reported by
\citet{Boehler2017ApJ...840...60B} and
\citet{Muto2015PASJ...67..122M}. In the analysis of
\citet{Boehler2017ApJ...840...60B} of CO isotopologue data, at the
position of the peak in dust continuum, the gas-to-dust mass ratio is
only $f_\mathrm{gd}=1.7$, increasing to $f_\mathrm{gd}\sim 20$ in the
direction of the minimum in continuum emission.  For the northern
shadow, there is general agreement in the literature on the dust
surface density peak, around $\Sigma_d \sim 1\,$g\,cm$^{-2}$. Since
$\Sigma = f_\mathrm{gd} \Sigma_d$, we vary only $f_\mathrm{gd}$. The
maximum grain size in the dust trap is at least $a_\mathrm{max} \sim
1\,$cm \citep[][]{Casassus2015ApJ...812..126C}, for which $\kappa_R =
0.05$\,cm$^2$\,g$^{-1}$ if $f=1$, and $\kappa_R =
0.32$\,cm$^2$\,g$^{-1}$ if $f=0.1$, so that $\tau_R \sim 4$ to $27$.
The cooling timescale is $\Delta t_c \sim 100 \times (f_\mathrm{gd} /
100)$\,yr. The shadow crossing time is $\Delta t_K \sim$152\,yr, for a
stellar mass of 2.0\,M$_\odot$
\citep[][]{Mendigutia2014ApJ...790...21M}, and for a shadow width
$\phi_S \sim$40\,deg.  Thus, if $f_\mathrm{gd}\sim 100$ under the
northern shadow, we expect dust to cool but not down to the floor
temperature, since $\Delta t_c \sim \Delta t_K$. Since $L_D / H \sim
1/\sqrt{\tau_R} \lesssim 1$, the smoothing effect of radiative
diffusion should have little impact. This scenario appears to be
observed, given the temperature drop of $\sim$30\% reported by
\citet[][]{Casassus2015ApJ...812..126C}.

\begin{figure}
\begin{center}
  \includegraphics[width=\columnwidth,height=!]{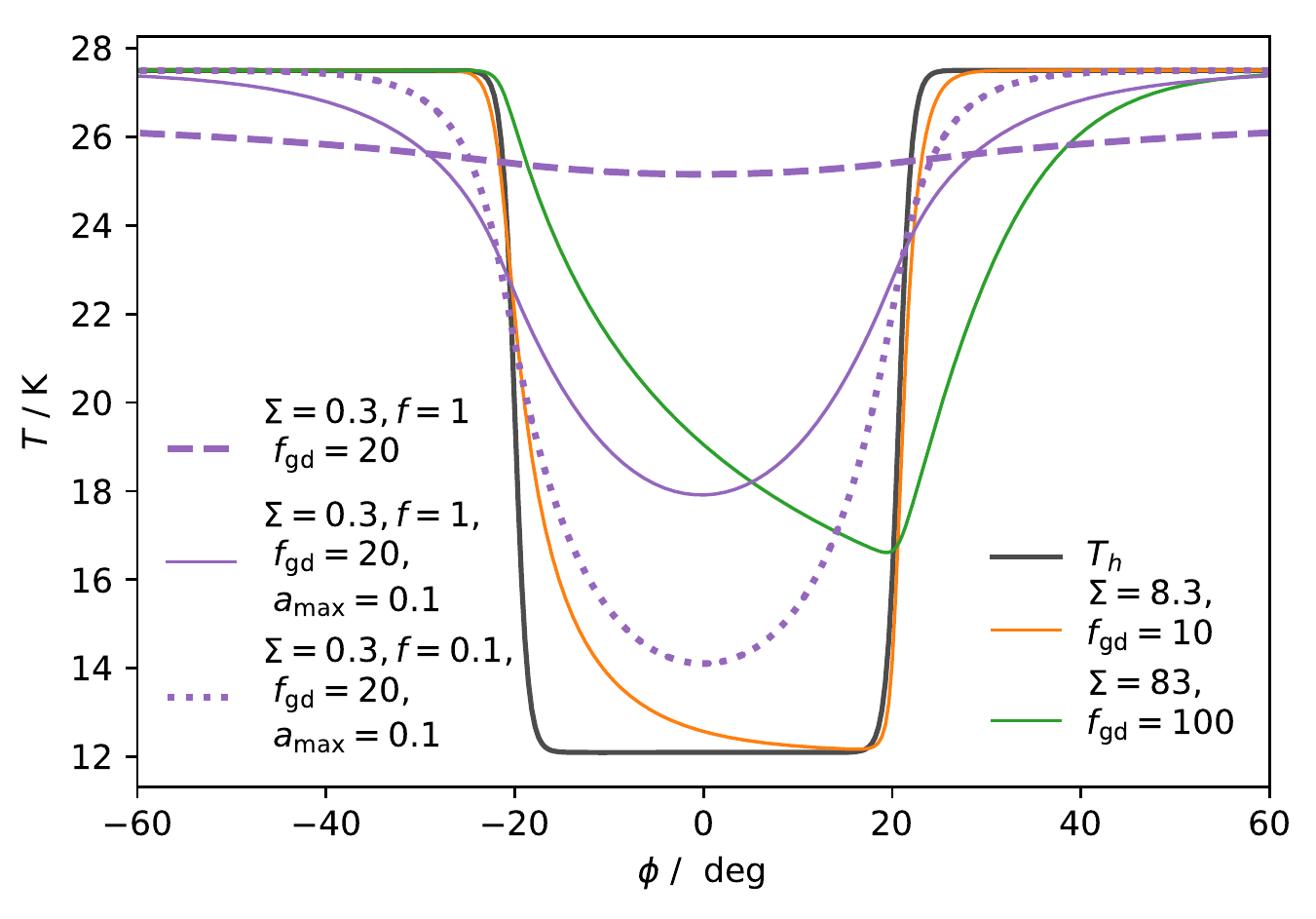}
\end{center}
\caption{Example profiles for $T(\phi)$ that approximate
  HD\,142527. All lines were computed for $a_\mathrm{max} = 1$\,cm and
  $f=0.1$ unless otherwise indicated in legends. The labels give
  surface density $\Sigma$ in g\,cm$^{-2}$. Gas flows towards
  $+\hat{\phi}$. \label{fig:HD142527}}
\end{figure}

Example solutions of Eq.\,\ref{eq:mastereq}, with the above
parameters, are shown in Fig.\,\ref{fig:HD142527}. We see that for
$f_\mathrm{gd} = 100$, radiative diffusion has no impact and the
profile follows a cooling curve close to Eq.\,\ref{eq:nodiffusion}. A
massive disk results in conspicuous shifts $\eta_S$ between the
temperature minimum and the centroid of the illumination pattern. For
$f_\mathrm{gd}=100$, $\eta_S = 19.1\,$deg if $f=0.1$ and
$a_\mathrm{max}=1\,$cm, while for $f_\mathrm{gd}=10$, $\eta_S =
14.4\,$deg if $f=1$ and $\eta_S = 16.2\,$deg if $f=0.1$.




For the southern shadow, in the RT$_\mathrm{1425}$ model the maximum
grain size that is not affected by azimuthal trapping is
$\sim$0.1\,cm, for which $\kappa_R = 0.35$cm$^{-2}$\,g$^{-1}$ if
$f=1$, and $\kappa_R = 1.1$cm$^{-2}$\,g$^{-1}$ if $f=0.1$ (both values
at 27.5\,K). We consider a single case for $\Sigma$: a gaseous disk
mass contrast of $ \sim$3 as in \citet{Boehler2017ApJ...840...60B},
with $f_\mathrm{gd} \sim 20$ and $\Sigma \sim 0.3$\,g\,cm$^{-2}$, so
$\tau_R = 1.65$ if $f=0.1$, and $\tau_R \sim 0.5 $ if $f=1$.  Gas
should cool much faster than in the northern shadow, since $\Delta t_c
/ \Delta t_K \gtrsim 10^{-2}$, but radiative diffusion should have
more of an impact. Example solutions are shown in
Fig.\,\ref{fig:HD142527}, where we also include a curve for
$a_\mathrm{max}=1$\,cm and $f=1$, for which $\tau_R \sim 0.015 $. In
this latter case the averaging effect of radiation transport almost
completely smooths out the temperature decrement.

%
%


\subsection{DoAr\,44}

\begin{figure}
\begin{center}
  \includegraphics[width=\columnwidth,height=!]{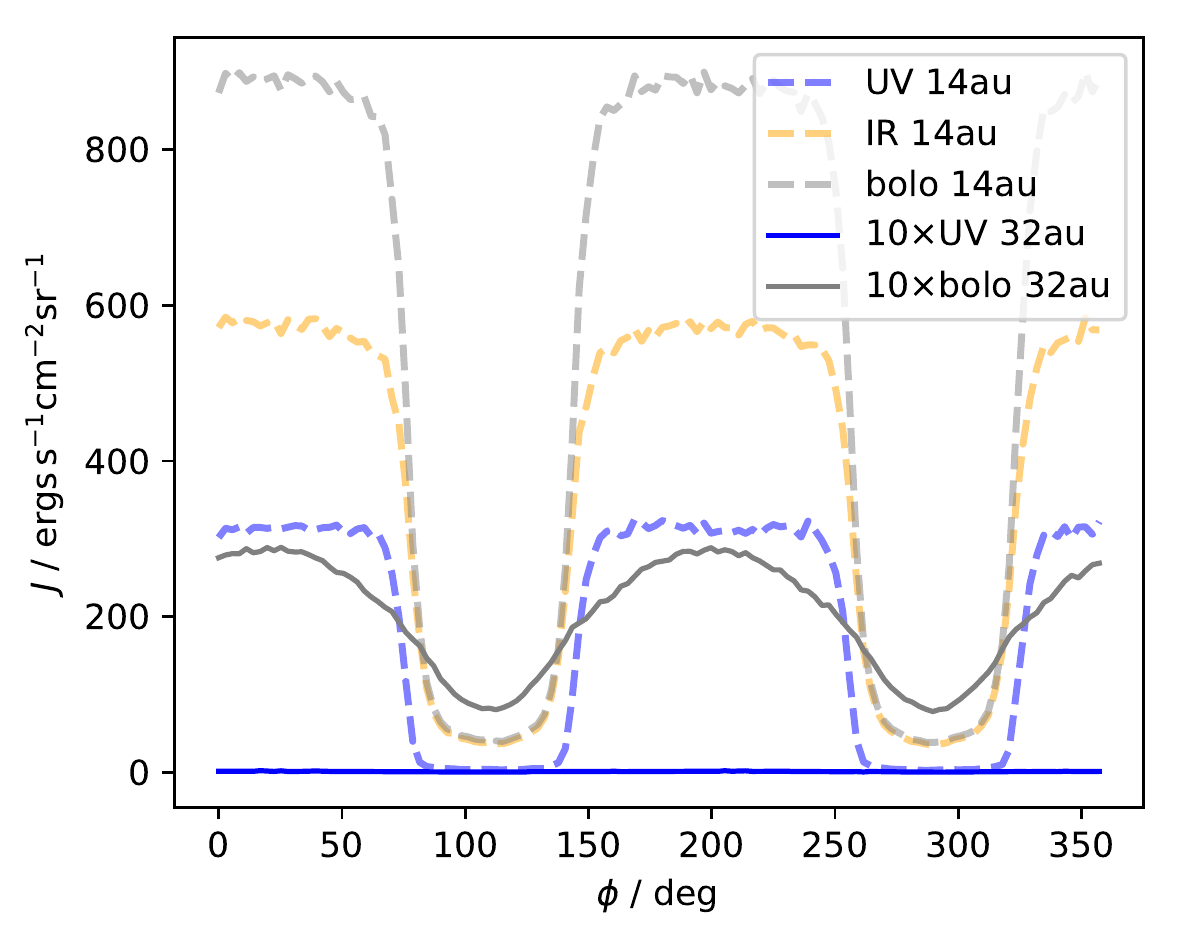}
\end{center}
\caption{Mean radiation intensity $J(\phi)$ in a RT model for
  DoAr\,44, without the outer disk. We plot $J(\phi)$ in the mid-plane
  and at two radii: at the edge of the gaseous cavity, so at
  $R_\mathrm{cavgas}=14$\,au, and at the edge of the large-dust
  cavity, at $R_\mathrm{cavdust}=32$\,au. The profiles at
  $R_\mathrm{cavdust}$ have been multiplied by 10. Conventions for the
  spectral domains follow from
  Fig.\,\ref{fig:J}.  \label{fig:DoAr44J}}.
\end{figure}

The cavity in DoAr\,44 is very different from HD\,142527. In the
parametric model of \citet[][ hereafter RT$_\mathrm{DoAr44}$
]{Casassus2018MNRAS.477.5104C}, the polarized-intensity image in
$H$-band is reproduced with a gaseous cavity radius
$R_\mathrm{cavgas}=$14\,au, while the continuum at 336\,GHz
corresponds to a larger cavity for the mm-sized dust, with
$R_\mathrm{cavdust}=$32\,au. Thus, if due to shadowing, the continuum
drops seen at 336\,GHz, of $\sim$24\%, result from a heat source
profile $T_h(\phi)$ that has been substantially processed by
intervening material, composed of gas and small dust (with radii
$a<1\mu$m in RT$_\mathrm{DoAr44}$). This is illustrated in
Fig.\,\ref{fig:DoAr44J}, which plots the bolometric mean intensity
field $J_\mathrm{bolo}$ at the edge of the gaseous cavity at
$R_\mathrm{cavgas}$ and at $R_\mathrm{cavdust}$, in the absence of the
outer disk. The broad and V-shaped decrements in $J_\mathrm{bolo}$ are
very different from square profiles.  Instead, for
$T_\mathrm{h}(\phi)$ we use $J_\mathrm{bol}^{1/4}$, extracted at the
edge of the large-dust cavity, i.e. at $R_\mathrm{cavdust} = 32~$au,
smoothed and normalized so that the peak temperature is that computed
in the RT$_\mathrm{DoAr44}$ model (with the outer disk, and for the
large dust population). The stellar mass for DoAr\,44 is
1.3\,M$_\odot$ \citep{Espaillat2010ApJ...717..441E}, and its distance
in RT$_\mathrm{DoAr44}$ is 120\,pc \citep[now revised to
  145.9\,pc,][]{Gaia2016A&A...595A...1G}. The RT$_\mathrm{DoAr\,44}$
model suggests that the disk aspect ratio is $h \sim 0.08 $ at
$R_\mathrm{cavdust}$, coincident with the peak gas surface density of
$\Sigma = 13$\,g\,cm$^{-2}$ for $f_\mathrm{gd} = 100$, if
$a_\mathrm{max}=1\,$mm.


\begin{figure}
\begin{center}
  \includegraphics[width=\columnwidth,height=!]{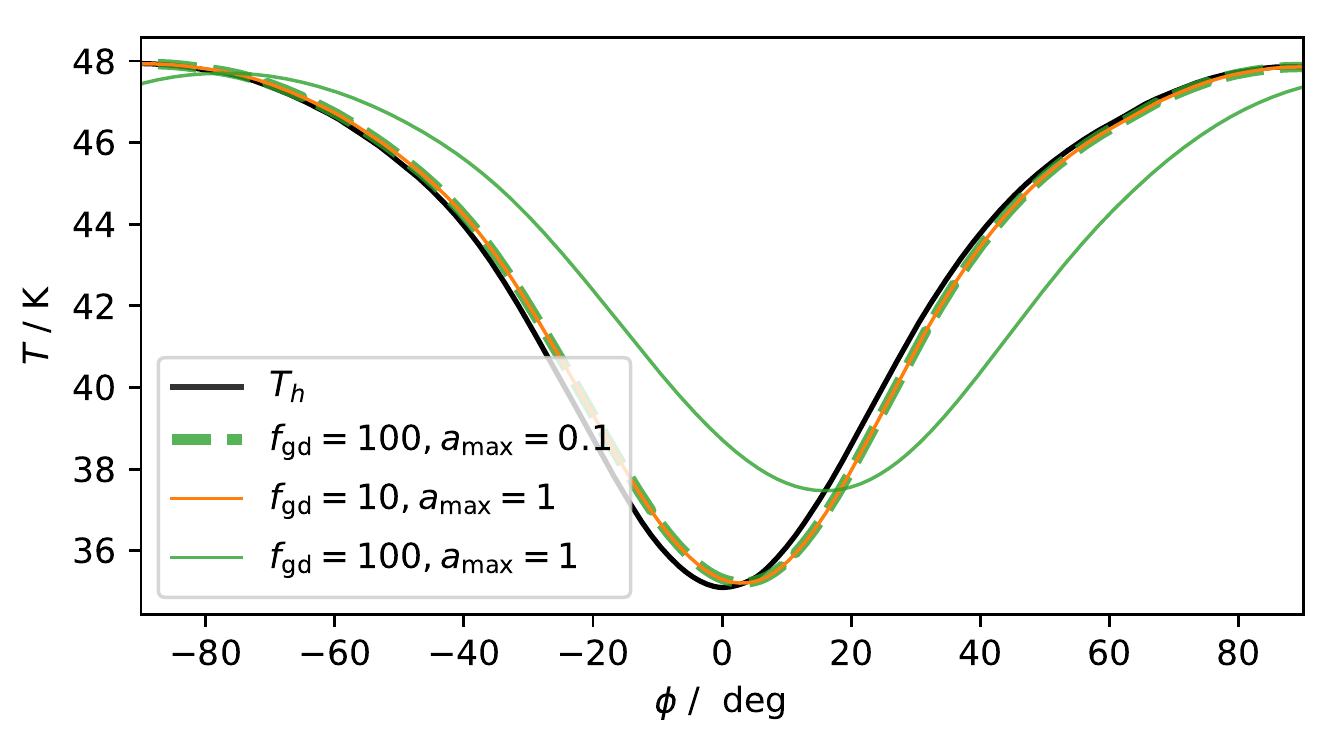}
\end{center}
\caption{ Example profiles $T(\phi)$ that approximate DoAr\,44. All
  curves correspond to $f = 1$, and we vary the gas-to-dust ratio
  $f_\mathrm{gd}$, for $a_\mathrm{max}=0.1$\,cm and
  $a_\mathrm{max}=1$\,cm. The heat source $T_\mathrm{h}(\phi)$ is
  shown in black line. The direction of Keplerian rotation is towards
  $+\hat{\phi}$.  \label{fig:DoAr44}}.
\end{figure}

As summarised in Fig.\,\ref{fig:DoAr44}, the effect of advection is
more important for a massive outer disk. In particular, the case for
$f_\mathrm{gd}=100$, $f=1$, $a_\mathrm{max} = 1\,$cm, with
$\kappa_R(48\,$K)=0.06\,cm$^{-2}$g, corresponds to $\Sigma =
130$\,g\,cm$^{-2}$ for the same mass in dust smaller than 1\,mm as in
RT$_\mathrm{DoAr\,44}$, and results in a large shift between the
azimuthal position of the minimum in $T_\mathrm{h}$ and that of the
minimum in $T$: $\eta_S = 16\,$deg. The curves for
$f_\mathrm{gd}=100$, $a_\mathrm{max} = 0.1\,$cm, and
$f_\mathrm{gd}=10$, $a_\mathrm{max} = 1\,$cm, both correspond to
$\Sigma = 13$\,g\,cm$^{-2}$ (given the grain size exponent of
$q=-3.0$), and $\eta_S = 2.9\,$deg. A run for $f_\mathrm{gd}=1$,
$a_\mathrm{max} = 1\,$cm ($\kappa_R(48\,$K)=0.40\,cm$^{-2}$g, not
shown in Fig.\,\ref{fig:DoAr44}), results in $\eta_S = 0.36\,$deg. In
all runs $\tau_R\gg 1$.

From the available observations \citet{Casassus2018MNRAS.477.5104C}
report $\eta_S = -2.8\pm0.9\,$deg, which is a shift along the
direction of rotation if the disk is rotating clock-wise (so if the
northern side is the far side). With this direction of rotation the
faint arc to the North-West \citep[seen in polarized
  intensity,][]{Avenhaus2018ApJ...863...44A} can be seen as a trailing
spiral arm. However, the substantial disk thickness in the $H$-band,
and the finite inclination, result in a model $\eta_s^m =+3.0$\,deg,
as calculated with the RT$_\mathrm{DoAr\,44}$ predictions at native
angular resolutions. After $uv-$plane filtering, the shift is
$\eta_s^m =-1.3$\,deg. If we take this offset of 4.3\,deg as an
indication of systematics, the observed value would be
$+1.5\pm0.9$\,deg. A very massive disk can thus easily be ruled out,
and we can also discard at 5$\sigma$ the cases with $f_\mathrm{gd} =
10$, $a_\mathrm{max} = 1$\,cm, and $f_\mathrm{gd} = 100$,
$a_\mathrm{max} = 0.1$\,cm. A 3D RT model that incorporates advection,
along with finer angular resolution data, could eventually refine
these limits.

%

\section{Conclusion} \label{sec:conclusion}




The midplane gas and dust cool when flowing across shadows cast by
inner warps. In this letter we propose a simplified 1D model to
explain the broad variety of temperature responses under the shadows,
and to assess how the shape and depth of the temperature profiles
depend on the outer disk mass.  At low surface densities $\Sigma$,
such that $\tau_R = \kappa_R \Sigma \ll 1$, or in the disk surface,
radiation will smooth out the temperature decrements.  In massive
rings, such that $\tau_R \gtrsim 1$, the temperature minimum will be
shifted in the direction of rotation relative to the centroid of the
illumination pattern due to advection.  This shift $\eta_S$ is a probe
of the gas-to-dust mass ratio for a given dust population.

Detailed applications to specific objects require the incorporation of
advection in 3D radiative transfer. Meanwhile we report some of the
prediction from our 1D model. The square illumination profile in
HD\,142527 allows to illustrate the impact of advection, even if in
the isothermal approximation. A massive disk with $f_\mathrm{gd} > 10$
near the continuum peak should result in a conspicuous shift,
$14\,$deg$ \lesssim \eta_S \lesssim 20 \,$deg. A comparison with
observations requires high angular resolution continuum data. In
DoAr\,44, the available observations rule out a massive
disk with $\Sigma > 13\,$g\,cm$^{-2}$ at 5$\sigma$, and given a
standard dust population.


\section*{Acknowledgments}

The referee, Kees Dullemond, provided important and constructive input
on the disk optical depth and the limits of the Rosseland
approximation. We acknowledge further useful comments from Zhaohuan
Zhu, Wladimir Lyra and Philipp Weber. Support was provided by
Millennium Nucleus RC130007 (Chilean Ministry of Economy), FONDECYT
grants 1171624 and 1151512, and by CONICYT-Gemini grant 32130007. This
work used the Brelka cluster (FONDEQUIP project EQM140101) hosted at
DAS/U. de Chile.

\bibliographystyle{mnras}

\input{mn_cool.bbl}



\bsp	
\label{lastpage}
\end{document}